\pdfoutput=1
\RequirePackage{ifpdf}
\documentclass[cits,11pt,a4paper]{article}
\usepackage{jheppub}
\usepackage{float}
\usepackage{caption}
\usepackage{subcaption}
\usepackage{siunitx}
\usepackage{graphicx}
\usepackage{booktabs}
\usepackage{natbib}
\usepackage{hyperref}
\usepackage{lineno}
\usepackage{amsmath}
\renewcommand{\theequation}{\arabic{equation}}
\usepackage{chngcntr}
\counterwithout{equation}{section}
\title{\boldmath Ion Transport on Phased Radiofrequency Carpets in Xenon Gas}

\abstract{We present the design and performance of a four-phased radiofrequency (RF) carpet system for ion transport in high-pressure xenon gas. The RF carpet, designed with a 160~$\mu$m pitch, is applied to the lateral collection of ions in xenon at pressures up to 600~mbar. We demonstrate transport efficiency of caesium ions across varying pressures, and compare with microscopic simulations made in the SIMION package. The novel use of an N-phased RF carpet at high pressure can achieve ion levitation and controlled lateral motion in a denser environment than is typical for RF ion transport in gases. This feature makes such carpets strong candidates for ion transport to single ion sensors envisaged for future neutrinoless double-beta decay experiments in xenon gas.}

\begin{document}

\author[1,a]{E.~Dey\note[a]{Corresponding author},}
\author[1]{B.J.P.~Jones,}
\author[1]{Y.~Mei,}
\author[2]{M.~Brodeur,}
\author[1]{V.A.~Chirayath,}
\author[1]{N.~Coward,}
\author[3]{F.W.~Foss,}
\author[1]{K.E.~Navarro,}
\author[1]{I.~Parmaksiz,}
\author[4]{C.~Adams,}
% ...
\affiliation[1]{
Department of Physics, University of Texas at Arlington, Arlington, TX 76019, USA}
\affiliation[2]{
Department of Physics and Astronomy, University of Notre Dame, Notre Dame, IN 46556, USA}
\affiliation[3]{
Department of Chemistry and Biochemistry, University of Texas at Arlington, Arlington, TX 76019, USA}
\affiliation[4]{
Argonne National Laboratory, Argonne, IL 60439, USA}
% ...

\flushbottom
\renewcommand{\theequation}{\arabic{equation}}
\maketitle
\section{Introduction}
\label{sec:intro}
 
Transport and trapping of ions in radiofrequency (RF) electric fields is a method with wide use cases in atomic and nuclear physics and analytic chemistry. The interaction between oscillating electric fields, charged particle motion, and interactions with buffer gases offers the opportunity to develop a myriad of schemes to realize ion transport, cooling, manipulation and trapping for applications with highly disparate requirements. Among other applications, RF-based structures have become commonplace in mass spectrometers~\cite{Mass_spec1997,Masssepec2} and gas-phase stopper cells in nuclear physics ~\cite{Riken,Caribu}.  At lower pressures, Paul traps~\cite{brown1991quantum} are ubiquitous tools for the precision study of ions for fundamental physics~\cite{horvath1997fundamental}.  RF manipulation and trapping also underlie the majority of trapped ion qubit designs~\cite{mehta2016integrated}, one possible path toward long-coherence-time quantum computing~\cite{wang2017single}. More recently, RF methods have been proposed for ion transport and/or extraction in ultra-rare decay searches, which is the application motivating the present work~\cite{brunner2015rf,jones2022dynamics}.

The NEXT collaboration aims to detect neutrinoless double beta decay ($0\nu\beta\beta$) in high pressure xenon gas. Detection of $0\nu\beta\beta$ is the most sensitive known method currently available to test whether neutrinos are Majorana particles, which could offer profound insight into the dominance of matter over antimatter in the Universe ~\cite{buchmuller2005leptogenesis}.  A key challenge in such experiments is the mitigation of backgrounds, primarily arising from radiogenic activity in detector materials and cosmogenic activations~\cite{dolinski2019neutrinoless}. To surpass current sensitivity limits, next-generation experiments must achieve ultra-low background levels ($b< 0.1 \, \mathrm{counts \, per \, ROI \, ton \, year}$)~\cite{agostini2017discovery} and deploy ton-scale masses of active isotope, requiring significant advancements in detector technology. Identifying or “tagging” the $^{136}$Ba daughter ion~\cite{Moe:1991ik} produced from the double-$\beta$ decay of $^{136}$Xe in xenon experiments would eliminate all backgrounds except for the double beta decay with neutrino emission $(2 \nu \beta \beta)$ which, in experiments with fine energy resolutions such as NEXT, is completely negligible~\cite{NEXT:2020amj}. Various methods have been proposed within the double beta decay community to detect the barium daughter~\cite{Chambers:2018srx,mong:2014iya,rollin:2011gla,sinclair:2011zz,flatt:2007aa,mcdonald2018demonstration,rivilla2020fluorescent,jones2016single,thapa2021demonstration,thapa2019barium,Bainglass:2018odn,herrero2022ba}.
The NEXT collaboration is pursuing the use of single-molecule fluorescence imaging (SMFI) to detect Ba$^{2+}$ in xenon gas~\cite{nygren2015detecting,jones2016single}. Prior to imaging at a sensor surface, manipulating or capturing the ion within dense xenon remains a significant challenge. The nEXO collaboration has pursued the use of an RF funnel extraction method to transfer ions from high pressure gases to vacuum with gas flow through an aperture~\cite{brunner2015rf}, and has discussed elaborations of this method for a scheme to extract ions from liquid xenon to a Paul trap~\cite{ray2024ion} for imaging via atomic fluorescence~\cite{flatt:2007aa}. 

The proposed Ba$^{2+}$ detection approach for NEXT, on the other hand, is to detect individual Ba$^{2+}$ ions at sensor surfaces via chelation with organic molecular reporters, a process that can take place directly within the xenon gas environment. Imaging of individual Ba$^{2+}$ ions in 10 bar xenon has recently been demonstrated by NEXT~\cite{Microscope}, suggesting that in-situ ion transport methods without gas flow will be preferable for ion concentration to the sensors, under this scheme. This leads to some simplification relative to vacuum-extraction approaches, as well as to additional complications. In particular, since there is no forcing effect of the gas flow to the imaging region, ions must be preserved and manipulated to bring them to SMFI sensors in stationary gas from the regions in the detector where they are produced. Due to the TPC drift field, Ba$^{2+}$ ions in NEXT detectors will be driven towards the cathode. They must then be transported laterally to fluorescence sensors without changing its charge state.  One approach to this problem is the use of RF carpets~\cite{jones2022dynamics}, which are interdigitated arrays of RF electrodes that can levitate and sweep ions across flat surfaces in the presence of buffer gases. An array of these RF carpets each  concentrating ions to a small region equipped with single molecule sensor can be used to cover the cathode cross-section, or they can be separated by electrostatic concentrators. RF carpets may be amenable to the task of ion concentration for $0\nu\beta\beta$ searches if and only if they can be be operated far outside their past regime of applicability in both pressure and voltage, to meet the needs of the high pressure environment. 

The RF transport dynamics on RF carpets can be described in terms of an emergent Dehmelt pseudo-potential.  According to Dehmelt~\cite{Dehmelt}, an ion with charge $q$ and mass $m$, placed in a spatially inhomogeneous electric field described as $\mathbf{E_0} (\mathbf{r}) \text{cos}(\Omega t)$, undergoes a micro-motion. Due to the spatial inhomogeneity and the fast-changing nature of the electric field, the ion experiences different field strengths during the period of one micro-cycle at the RF frequency $\Omega$. As a result, even though the time integrated electric field at every position in space is zero, due to the correlation between the micro-motion and the applied electric field, the ion feels a net force that can be described using a pseudo-potential $V$~\cite{Schwartz},
\begin{equation} 
    V = \frac{q}{4m(D^2 + \Omega^2)} E_0^2 (\mathbf{r}).
    \label{eq:1}
\end{equation}
Here $D = \frac{q}{\mu m}$ is the damping factor introduced by collisions with the buffer gas, and $\mu$ is the ion mobility. For an RF carpet operated based on a purely N-phased RF voltage, the voltage on a $j$th electrode at time $t$ for an RF peak-to-peak voltage of $V_{\text{pp}}$ is given by

\begin{equation}
    V_j (t) = \frac{V_{\text{pp}}}{2}\sin \left(\Omega t + \frac{2\pi j}{N}\right).
    \label{eq:2}
\end{equation}
and as shown in Ref~\cite{jones2022dynamics}, the circular micro-motion in this case leads to a repulsive pseudo-potential, which is superimposed on a direct-current (DC) electric push field $E_{push}$ towards the carpet,
\begin{equation}
    V= \frac{q^2}{m(D^2 + \Tilde{\Omega}^2)}\frac{1}{2}\left(\frac{2\pi}{Np}\right)^2 \left(\frac{V_{\text{pp}}}{2}\right)^2  \text{exp} \left(\frac{-4 \pi}{Np}y\right) +qE_{push} y,
    \label{eq:3}
\end{equation}
 where $p$ is the carpet pitch and y is the vertical distance from the carpet. This resultant force can be utilized to trap ions above the surface of RF carpets geometries.  Unlike in a simple Paul trap with a stationary attractor, in this system the relevant frequency of the micro-motion is not exactly equal to the drive frequency $\Omega$, because the ion is moving laterally while passing through the RF field. The modified frequency driving the micro-motion is given by $\Tilde{\Omega}$
\begin{equation}
    \Tilde{\Omega} = \left(\Omega+ \frac{2\pi}{Np}\left<\mathbf{v}(y)\right>\right)
    \label{eq:4}
\end{equation} where  $\left<\mathbf{v}(y)\right>$ is the mean ion velocity at fixed height y above the carpet surface. The  number of RF phases $N$ applied between the adjacent electrodes determines the form of the micro-motion and the emergent macro-motion.

Calculations in two-phase (each adjacent electrode carrying 180$^\circ$ phase differences) RF carpet geometries~\cite{Schwartz} show that the emergent force from the pseudo-potential is perpendicular to the carpet surface. It is then possible to engineer conditions such that the ion travels transverseley in one preferential direction, and these schemes fall into a variety of classes.   One especially notable approach is the ``ion surfing'' technique, proposed by Bollen~\cite{Bollen2011}. In this method, RF signals are applied to adjacent electrodes with a 180$^\circ$ phase difference to get the repulsive force through the Dehmelt potential, while a superimposed lower frequency and lower voltage traveling wave sweeps the ions laterally.  This approach has been successfully implemented in various experiments~\cite{MWada_circ_RFC,Maxime2022RFC,RFC3,RFC4,RFC5,Lund2020}, typically in pressures of up to 100~mbar in helium buffer gas.

Emergent Dehmelt dynamics are expected whenever the mean-free path between buffer gas collisions is a significant fraction of one micro-cycle. In these conditions, the ions undergo accelerated or curved rather than effectively straight trajectories between collisions, and it is this acceleration or curvature that yields an E-field-to-position correlation to generate the pseudo-potential. As the gas pressure increases, collisions with buffer gas atoms limit the micro-motion resulting in a decrease in the RF repulsive force as experienced by the ion. Maintaining ion levitating dynamics at higher pressures therefore requires smaller electrode pitches, higher RF voltages, and higher RF frequencies.
These changes come at a cost of equilibrium ion trajectories that sit closer to the electrode surface.  Our previous work found that in the conditions expected for ion transport in modest-to-high pressure (1-10~bar) xenon gas, this near-surface transport makes the Bollen ion-surfing method unfeasible, as long transport distances are difficult to achieve given the chaotic two-frequency motion.  

Another approach to lateral transport is to superpose a transverse DC bias on top of the RF levitating voltages.  This method has the advantage of being implementable with capacitive and resistive elements in a PCB-based carpet and yielding a simple, analytic form for the combined micro and macro-motion, but it becomes impracticable over appreciably large transport distances due to the necessarily large potential difference that must be applied between the two sides of the carpet.  

A third method of achieving lateral ion transport is to use $N$ phases at a single RF frequency~\cite{jones2022dynamics}, a scheme that has been advanced by the NEXT collaboration and appears to be especially well suited to high pressure environments. When $N\geq3$, the emergent ``spiral'' micro-motion leads to both a levitating and transverse sweeping force, and adequate ion transport performance can, in principle, be maintained in realistic structures up to several atmospheres of pressure.   To obtain transport in the highest possible pressures, devices should be manufactured with the smallest possible pitch and operated at the highest possible voltage with frequencies in the MHz regime. The maximum RF voltage is constrained by the breakdown threshold of either the gas or the insulating material of the carpet, whichever is limiting. We have found that for our devices in xenon gas at all relevant pressures, it is the carpet substrate rather than the gas that undergoes breakdown at higher RF amplitudes. 
 
In an $N$-phased system, the maximum voltage difference between adjacent electrodes $V_{\text{max}}$, with applied RF voltage of $V_{\text{pp}}$ is related as
\begin{equation}
V_{\text{max}} = V_{\text{pp}} \sin\left(\frac{\pi}{N}\right).
\label{eq:5}
\end{equation}
For $N>2$, the voltage experienced between adjacent electrodes is reduced compared to the RF peak-to-peak voltage. Analytical calculations suggest that a soft optimum between levitating action and breakdown resilience exists for $N\sim4-6$. Of these options, an $N=4$ phased case presents the least RF engineering challenges, and appears to be a suitable design for balancing the challenging transport dynamics of ions in high pressure buffer gases.

In this paper, we demonstrate the application of a four-phased RF carpet in xenon gas across various pressure conditions. We drive the carpet with a $\sim$MHz RF wave that is phase-delayed by $90^\circ$ between each pair of adjacent electrodes. Section~\ref{sec:App} describes the RF carpet geometry, RF drive circuit, ion source used in the experiment, and the test setup, including the carpet and electrodes arrangement inside the pressurized vessel. Section~\ref{sec:Measure} outlines the methodology for measuring the ion current at the collection region. Section~\ref{sec:result} discusses the results obtained in this experiment, including the transport efficiency measured at room temperature in xenon gas at pressures between 200~mbar and 600~mbar, and presents a comparison to theoretical predictions of transport efficiency and their qualitative and quantitative precision. Finally, Section~\ref{sec:conc} provides the conclusions of this study and outlines future directions for improving RF carpet performance in high-pressure environments.

\section{Apparatus}
\label{sec:App}

This section outlines the main elements of the experimental apparatus. At the core of the system is a 160~$\mu$m pitch RF carpet of thickness 0.2~mm operating in a four-phase configuration at moderate pressures in xenon gas. This RF carpet represents the state-of-the-art in printed circuit board (PCB) manufacturing for ion transport applications, as detailed in Section~\ref{sec:Carpet}. The implementation of the phased-driven mode necessitated the development of a custom high-voltage RF power circuit capable of delivering a four-phased signal, as described in Section~\ref{sec:Circuit}. Ion generation was performed using an aluminosilicate ion source, selected for its stability and compatibility with xenon gas environments, which is discussed in Section~\ref{sec:Source}. Finally, Section~\ref{sec:Exp} explains how these components were integrated into the experimental setup, providing a comprehensive configuration for testing ion transport efficiency.

\begin{figure}[t]
\begin{centering}
\includegraphics[width=1\columnwidth]{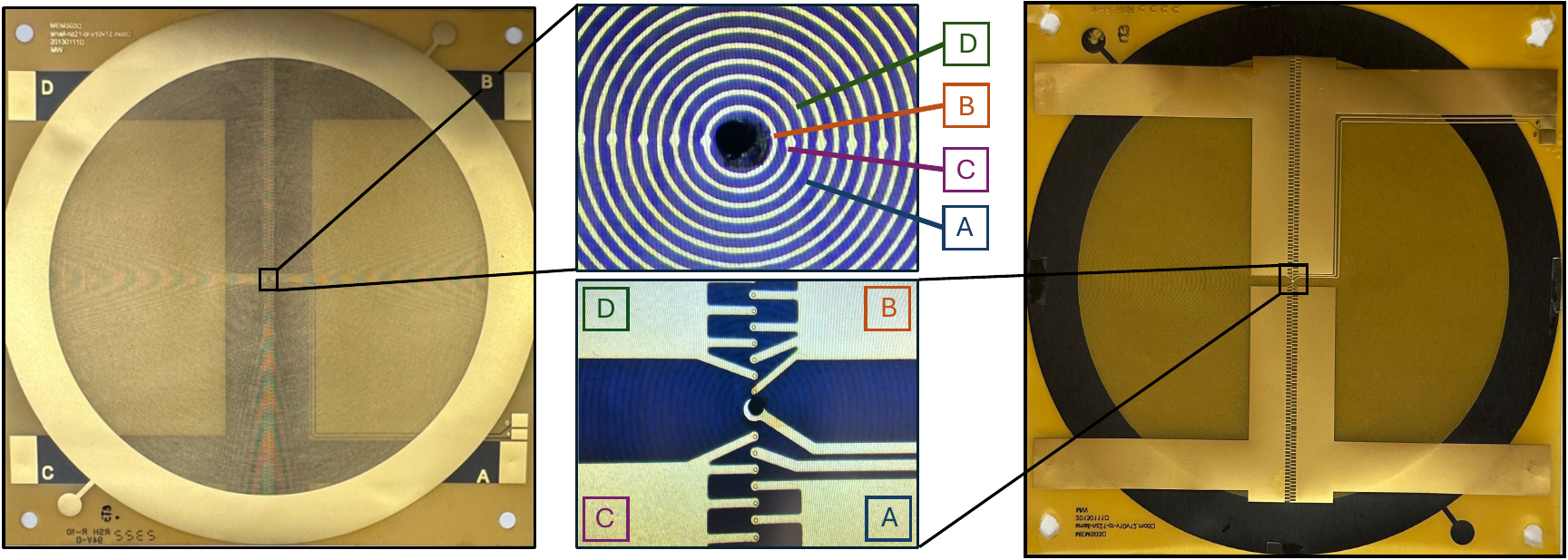}
\par\end{centering}
\caption{(Left) RF Carpet used for transporting Cs$^+$ ion in Xe gas (200~mbar to 600~mbar). (Right) Back side of the RF Carpet. (Center top) magnified top view of the center of the RF carpet, and (Center bottom) magnified bottom view of the center of the RF carpet that shows how the concentric electrodes are connected to four pads A, B, C, and D through PCB vias, resulting in a four-phase RF carpet.\label{fig:Carpet}}
\end{figure}

\subsection{Carpet\label{sec:Carpet}}
RF carpets were originally introduced by M.~Wada at RIKEN~\cite{MWada_circ_RFC} for use in nuclear physics stopping cells.  Three manufacturing methods are prevalent. Carpets can be produced by stretching a wire array across free space~\cite{ringle2021particle}, or by assembling stacks of thin electrodes in a cone configuration, an especially convenient geometry for use in gas catchers~\cite{savard2016caribu,Masssepec2}.  The smallest pitch RF carpets to date are made via precision PCB manufacturing~\cite{MWada_circ_RFC}.

The PCB-based carpet used in this work (Fig.~\ref{fig:Carpet}) was designed for the Superallowed Transition BEta NEutrino Decay Ion Coincidence Trap (St. Benedict) experiment ~\cite{Maxime2022RFC,St.Benedict} and was manufactured by RushPCB, for operation in ion-surfing mode~\cite{Bollen2011}. The carpet features a central hole surrounded by 264 concentric ring electrodes. Each electrode is connected by a through-hole (via) to the backside of the board. Every fourth electrode is linked to a distinct strip on the back through open vias, forming four groups (each having 66 electrodes) of electrodes labeled A, B, C, and D. Surrounding the carpet electrodes is a 1-cm-wide ring electrode, referred to here as the “collection ring", which serves as the region where ion transport efficiency is measured. The entire PCB measures $10\times10$~cm$^2$. The center-to-center distance between adjacent electrodes (pitch) is 160~$\mu$m, while the distance from the edge of the last electrode to the inner boundary of the collection ring is 120~$\mu$m. Our use case for the carpet is for collection rather than through-hole extraction, and due to the difficulty of instrumenting the central electrode with a collection circuit, we have tested the carpet in an outward-sweeping rather than inward-sweeping mode---collecting ions on the outer ring. For future devices that will be made specifically for our application, we will modify the design to employ an inward-sweeping mode with a central collection electrode and our results can be reasonably expected to map to those conditions.

\subsection{RF drive electronics\label{sec:Circuit}}

As noted above, effective ion transport above a carpet surface necessitates the repulsion of ions from the surface and a directing force towards a collection region. This is achieved by applying four RF signals with a $90^\circ$ phase difference on every pair of adjacent electrodes. This phased RF wave not only repels ions but also facilitates their lateral transport across the carpet. To ensure stable ion levitation, a constant DC push field counterbalances the RF repelling force.

Supplying a four-phased, $\sim$2 MHz RF signal with a magnitude $V_{\text{pp}}\gtrsim300$~V to the carpet is a major technical challenge.  Typical RF signal generators and amplifiers are designed (or specified) for 50~$\Omega$ load.  $V_{\text{pp}}=300$~V output would correspond to a 225~W power, which is quite large.  However, an RF carpet is not a 50~$\Omega$ load.  Due to the construction---a series of concentric ring electrodes---the carpet is expected to be a primarily capacitive load. In a purely capacitive load, the current and voltage are out of phase, and the device itself does not dissipate significant real (resistive) power. Instead, most of the energy is exchanged back and forth between the source and the capacitive field. While some small losses may occur due to dielectric heating or imperfect conductors, these are generally expected to be negligible compared to resistive heating in a 50~$\Omega$ load. As a result, it is expected that the carpet will not dissipate an appreciable amount of power.

\begin{figure}[t]
  \centering
  \begin{minipage}[t]{0.45\textwidth}
    \centering
    \includegraphics[width=0.9\textwidth]{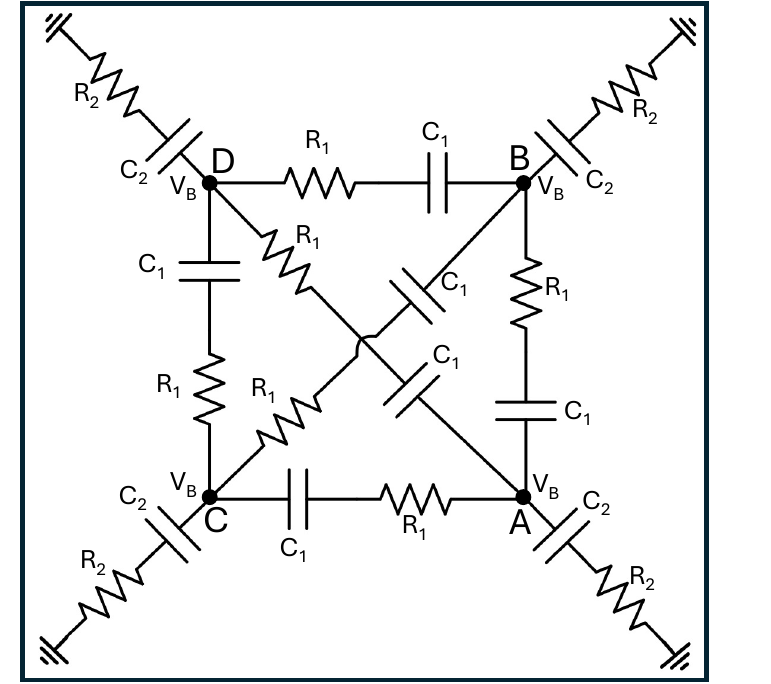}
    \caption{A lumped element model of the RF carpet. Labels A, B, C, and D indicate the phase nodes to which the RF drive signals connect. V$_\text{B}$ denotes the DC bias connection. The rest of the circuit elements represent equivalent capacitance and resistance (including RF radiative loss) of the carpet.}
    \label{fig:Carpet_circuit}
  \end{minipage}%
  \hfill
  \begin{minipage}[t]{0.45\textwidth}
    \vspace{-160pt} % Ensure alignment with the figure
    \begin{tabular}{cc}
        \toprule
        Parameters & Values \\ \midrule
        A$_1$, A$_2$ & HF linear amplifiers \\
        T$_1$, T$_2$ & Air-core center-\\ &tapped transformers \\
        N$_\text{P}$ & 18 turns \\
        N$_\text{S}$ & 36 turns \\
        V$_\text{B}$ & 16.8~V \\
        C$_1$ & 451~pF \\
        C$_2$ & 285~pF \\
        R$_1$ & 215~m$\Omega$ \\
        R$_2$ & 233~m$\Omega$ \\ \bottomrule
    \end{tabular}
    \vspace{0.4em} % Adjust space between table and caption
    \captionof{table}{Component values and electrical parameters relevant for the carpet model Fig. \ref{fig:Carpet_circuit} and the RF driving circuits Fig. \ref{fig:RFC_circuit}.}
    \label{tab:parameters}
  \end{minipage}
\end{figure}

We constructed a lumped element model to describe the behavior of the carpet as seen in the RF circuitry.  The model circuit is shown in Fig.~\ref{fig:Carpet_circuit}.  The capacitors in the circuit represent capacitances between phases and between phase electrodes and a ground plane on the carpet.  The resistors represent the resistive losses in the electrodes and the minor radiative loss of RF.  We used a vector network analyzer (VNA)~\cite{VNA} to measure the RF response of the carpet among the phases.  The component values in the lumped element model are determined from these measurements by matching with SPICE~\cite{spice} simulation.  They are shown in Tab.~\ref{tab:parameters}.  Note that all 4 phases are equal (symmetric) and mainly capacitive.  Within the frequency range of interest, the inductive elements are small and, therefore, ignored.

To match the output impedance of a typical RF amplifier, which is 50~$\Omega$, and the RF carpet, which is primarily capacitive (high impedance), we elected to use a transformer.  The transformer solution provides several advantages for this application:
\begin{itemize}
\item Impedance transformation.  For an ideal transformation, the impedance of the secondary winding is $n^2$ of that of the primary winding, where $n$ is the secondary-to-primary turn ratio.  A transformer with $n>1$ generally fits the requirement, which is matching a low (50 $\Omega$) impedance primary input to a high-impedance secondary output.
\item Voltage step up.  With $n>1$, the voltage of the secondary is higher than that of the primary.  This feature allows the output to reach the desired $V_{\text{pp}}\gtrsim300$~V while keeping the primary, which equals the output of the amplifier, to a lower voltage. As a result, a wide range of low-power RF amplifiers that cannot output high voltage can be used.
\item Galvanic isolation.  The RF carpet is to be biased at a DC voltage of $V_B$. The collection ring must operate at a lower voltage than the RF carpet but at a higher voltage than the grounded vessel. Consequently, the RF carpet needs to be maintained at a positive voltage higher than that of the collection ring to establish the required potential gradient for efficient ion extraction from the carpet to the collection ring. Because the secondary windings of the transformer are electrically isolated from the primary winding, the DC bias can be directly applied on the tap of the secondary windings and the secondary can have a different potential than the primary.  This tremendously simplifies the power delivery to the signal source and amplifiers and eliminates potential grounding issues.
\item Phase generation.  Using two secondary windings of opposite polarity, a single input can generate two output phases 180$^\circ$ apart.  To generate 4 phases 90$^\circ$ apart, only two RF drivers with 0$^\circ$ and 90$^\circ$ are needed.  The other two phases are generated by two transformers, which flip the 0$^\circ$ and 90$^\circ$ phases while simultaneously amplifying them.
\end{itemize}
\begin{figure}[t]
\begin{centering}
\includegraphics[width=1\columnwidth]{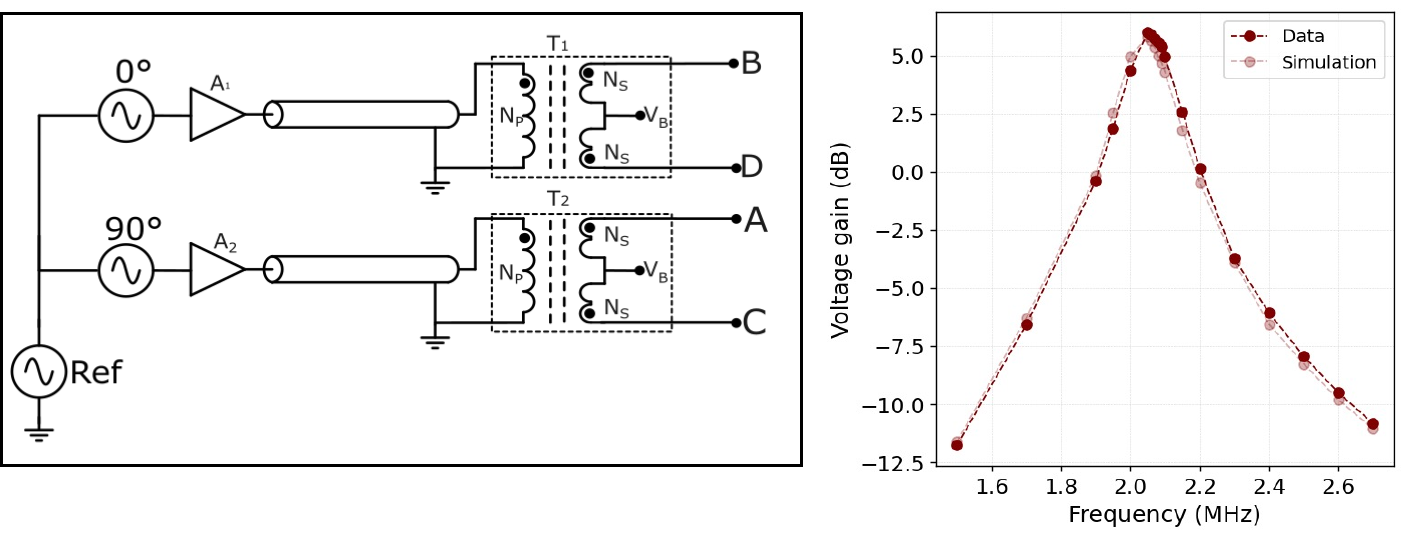}
\par\end{centering}
\caption{(Left) Block diagram of the RF drive signal chain.  RF power, superposed on top of a DC bias voltage $V_B$, is supplied to the electrodes A, B, C, and D of the carpet with the required phase relations. The circuit provides 90$^\circ$ phase difference between adjacent electrode pairs. The circuit parameters are listed in Tab.~\ref{tab:parameters}. (Right) The frequency response of the transformer-RF carpet system with resonance. The RF carpet is operated at the resonance frequency.\label{fig:RFC_circuit}}
\end{figure}

 As shown in Fig.~\ref{fig:RFC_circuit} (left), two signal generators generate the 0$^\circ$ and 90$^\circ$ phases.  The two signal generators are locked to the same reference signal; therefore, the two phases are coherent.  Each phase is amplified by a linear amplifier~\cite{Amplifier} before entering a coax cable.  At the other end of the coax cable, the RF signal enters the primary side of the transformer.  Each transformer has two secondary windings that share a terminal (tap).  The tap is connected to a DC bias voltage $V_B$.  The outputs of the two transformers, phases A, B, C, and D, are connected to the RF carpet through short wires.  Through experimentation, we elected to construct the transformer by winding 26~AWG Litz wires on a 3D-printed toroidal air core.  The turn ratio was selected to be $n=2$ (N$_\text{S}:$ N$_\text{P}$).  This construction eliminates magnetic core saturation and allows MHz RF operation.

 Due to the inductance of transformer windings $L_T$, the transformer and the carpet, form an LC circuit with certain resonance frequency.  We tuned the number of windings of the transformer such that the resulting $L_T$, together with the RF carpet, exhibit a resonance exactly at where the optimal driving frequency of the RF carpet should be.  Fig.~\ref{fig:RFC_circuit} (right) shows the measured frequency response of the transformer-RF carpet system.  At resonance, the gain approaches 6~dB at the carpet electrode from the amplifier output, which is close to the expected values of a factor of 2 set by the turn ratio.

 The tuning of the circuitry is guided by comparing measurements to SPICE simulation results.  The lumped element model of the carpet, the transformer's windings and coupling coefficient, the coax transmission lines, and the RF driver characteristics are all taken into account.  Fig.~\ref{fig:RFC_circuit} (right) demonstrates that the measurements and simulation are well matched.

\subsection{Ion source\label{sec:Source}}

An aluminosilicate ion source~\cite{Aluminosilicate_filament_source,Weber_Cordes_1966} was used for the transport test, manufactured by Heatwave Labs (Model 101139)~\cite{Cathode_TechBulletin_118}. Aluminosilicates are effective alkali ion sources, particularly in vacuum or noble gas environments, due to their stability and ion-exchange properties.  The aluminosilicate structure, with its large cavities, allows alkali ions (Na, K, Cs, etc) to migrate to the surface and be emitted as positive ions when sufficient thermal energy is applied. Thus, when sufficiently heated, aluminosilicate matrices release ions in a controlled manner in the 1+ charge state. Their porous nature and high thermal stability ensure sustained ion emission without significant degradation or reaction with ambient air.

\begin{figure}[t]
\begin{centering}
\includegraphics[width=1\columnwidth]{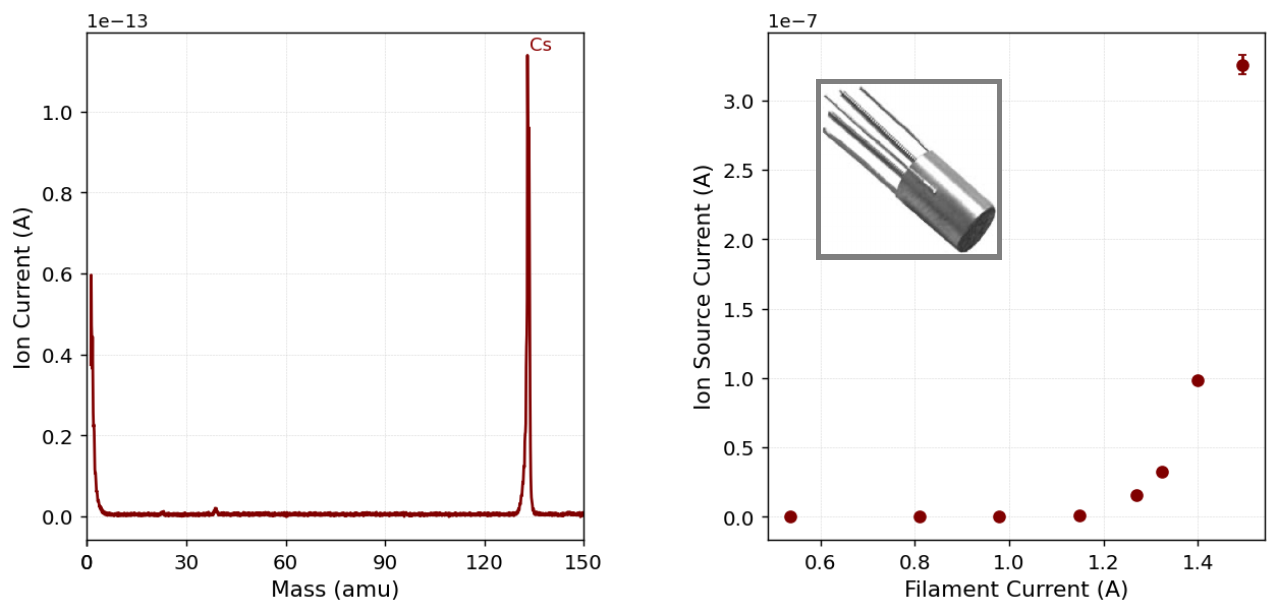}
\par\end{centering}
\caption{(Left) RGA mass spectrum confirming the production of pure Cs$^{+}$ ions from the ion source. (Right) Ion source output current as a function of the filament current used for heating. The inset image shows the ion source assembly itself.\label{fig:Cs source}}
\end{figure}

Since the primary objective of this experiment is to test RF carpets for barium tagging, a test ion with similar mass to barium was selected. Stability of ions in RF carpets at high pressure is dictated significantly by the mass ratio of transported ion to buffer gas, becoming more challenging as these masses become similar due to the large Brownian fluctuations introduced by collisions. A caesium ion source was thus selected for testing due to its proximity to the mass of barium. The ion source was mounted on an isolated copper feedthrough attached to a CF~2.75" vacuum flange. Our  simulations suggest that transport of dications will be enhanced relative to monocations in our conditions of interest, and as such, the results with Cs$^+$ can be considered as a conservative proxy for the expected transport efficiencies of Ba$^{2+}$.

Initially, the ion source was characterized in a standalone setup using a residual gas analyzer (RGA) to check for performance and ion yield in vacuum. The system was evacuated to a pressure below 10$^{-7}$ Torr. The current was gradually increased to 1.5 A, and a DC bias of 100 V was applied to accelerate the ions toward the RGA aperture. During this measurement, the RGA ionizer was turned off, so only ions arriving at the quadrupole with nonzero charge are recorded. Fig.~\ref{fig:Cs source} (left) presents the average of ten RGA scans, where the current peaks at a mass of 132.9~amu with negligible activity elsewhere, confirming the purity of Cs$^+$ ions. Since the ionization potential of caesium is around three times lower than that of xenon, no charge transfer from caesium to xenon is anticipated in these experiments. Furthermore, negligible ion current from Xe$^+$ is expected at temperatures near where the Cs$^+$ current begins to turn on, though notably this would also be a similarly valid ion to use to test the carpet transport properties. In the experiment depicted in Fig.~\ref{fig:Cs source} (right), a metal plate was used to measure the current with a picoammeter. The ion source was biased at 70 V, and the plate at 50 V. The source began producing stable ions at heating currents of approximately 1.3~A. 

\begin{figure}[t]
\begin{centering}
\includegraphics[width=0.8\columnwidth]{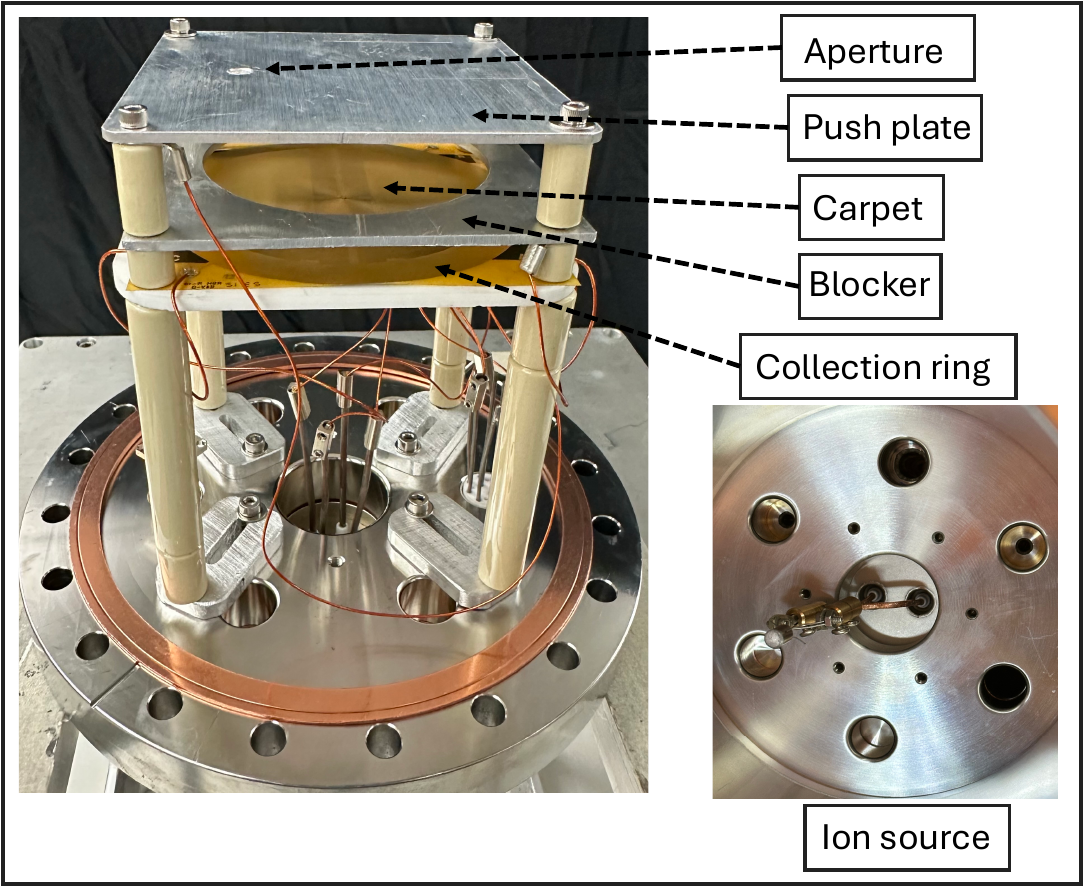}
\par\end{centering}
\caption{Picture of the experimental setup inside the pressurized vessel, including the caesium thermionic ion source and RF carpet test stand.\label{fig:RFC setup}}
\end{figure}

\begin{figure}[t]
\begin{centering}
\includegraphics[width=1\columnwidth]{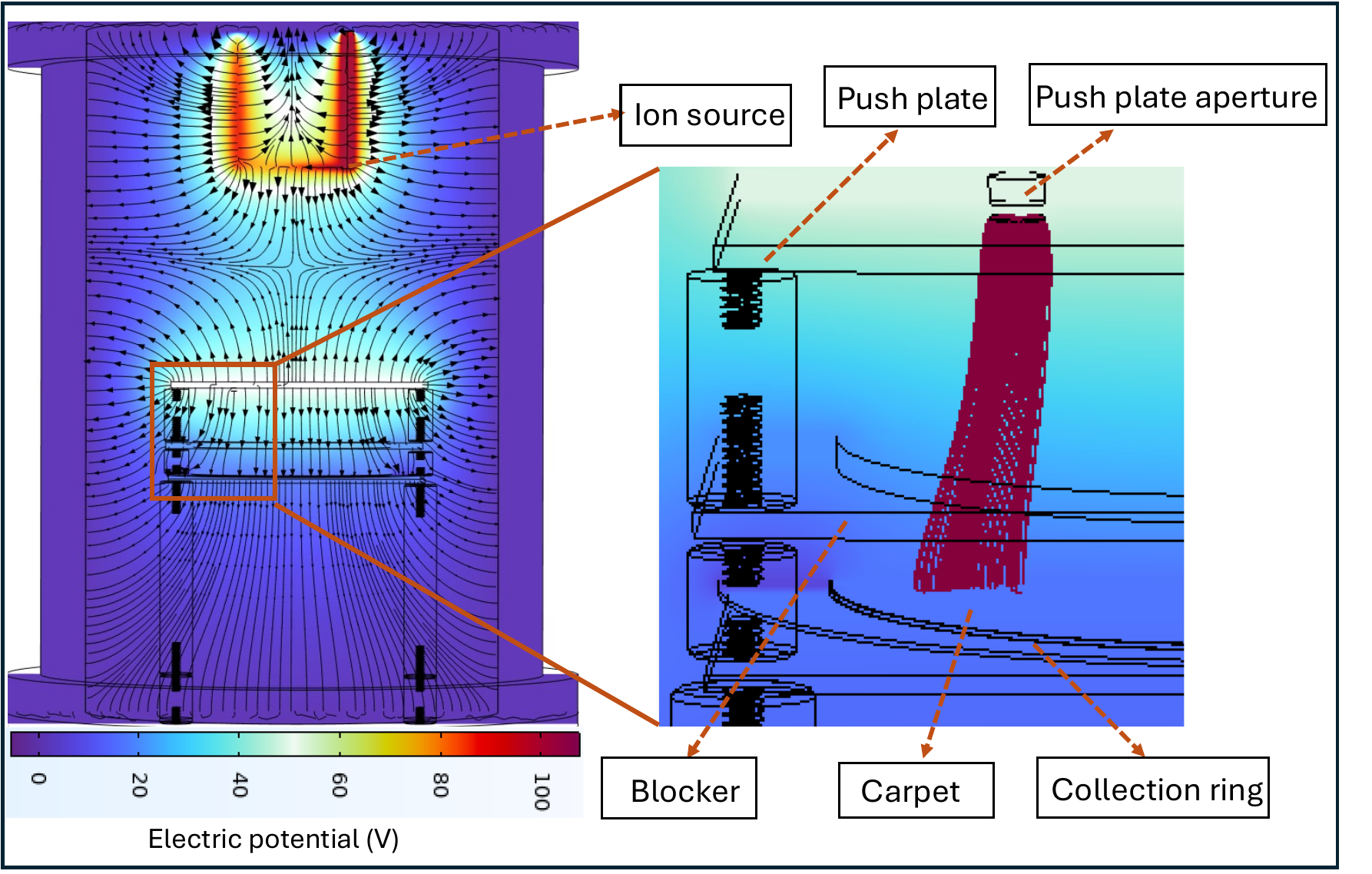}
\par\end{centering}
\caption{COMSOL simulation of the electric potential distribution inside the chamber. The color gradient represents the applied electric potential, while the lines and arrows depict the electric field direction. The zoomed-in view highlights the field lines originating from the push plate aperture and terminating on the RF carpet, with all other field lines hidden for clarity.\label{fig:Electricfield}}
\end{figure}

\subsection{Experiment configuration}\label{sec:Exp}

The experimental setup (Fig.~\ref{fig:RFC setup} (left)) consists of a caesium thermionic ion source mounted on one side of the vacuum vessel, while the opposite side houses the RF Carpet test stand. The ion source filament is connected to two copper feedthroughs on a CF flange, which serves both to heat the source (AC current) and to apply a positive bias to the ion source (DC voltage), with AC and DC components isolated via a high voltage transformer immersed in mineral oil. The carpet is supported on a 3~mm-thick sheet of macor insulator. A stainless steel push plate is positioned 35~mm in front of the carpet, separated by PEEK spacers. An electric field is established between the ion source and the push plate, creating a net force that accelerates ions towards the RF carpet. %pushes ions forward. 
The push plate has an aperture extending 5 mm and centered 15 mm from inner boundary of the collection ring. The electrodes on the carpet are biased to steer the ions through the aperture in the push plate towards the carpet surface roughly 15~mm away from the collection ring. The collection ring is maintained at a lower positive bias compared to the carpet, ensuring efficient extraction of ions from the carpet's final electrode to the collection ring. To prevent ions from directly reaching the collection ring, which is negatively biased relative to the carpet, a blocking electrode is placed between the push plate and the collection ring 12 mm away from the carpet. This electrode inhibits collection of ions that, primarily due to diffusive effects, would otherwise be deflected toward the collection ring.

To evaluate the behavior of the DC electrodes and gain insight into ion trajectories, a comprehensive internal electric field map was generated using the COMSOL Multiphysics simulation tool~\cite{COMSOL} (see Fig.~\ref{fig:Electricfield}). In these simulations, the vacuum chamber walls are at ground potential, causing most field lines originating from the source region to terminate at the chamber walls. Because the experimental configuration is designed to draw only a very small (pA) current (to avoid space charge effects) onto the carpet, nearly all ions emitted from the source land on the chamber's wall and upper side of the push plate, while a tiny fraction  of the ions pass through the aperture, as a result of diffusion superimposed on drift induced by the field. Fig.~\ref{fig:Electricfield} (zoomed-in) shows the field lines for trajectories emanating from this aperture to the carpet surface. These simulations confirm that the majority of electric field lines emerging through the push plate aperture indeed focus onto the carpet, thereby demonstrating the effectiveness of the electrode arrangement in guiding ions toward the intended collection area.

\section{Measurement}
\label{sec:Measure}

The system is initially evacuated to a pressure below 10$^{-7}$ Torr using a Pfeiffer HiCube 80 Classic Turbopump. To prevent oxidation or contamination when exposed to atmospheric air, the ion source is stored in an argon environment whenever the system is opened. Subsequently, the ion source is baked under vacuum conditions until outgassing ceases. The experimental setup is connected to a xenon gas cylinder equipped with both inlet and outlet lines, and a circulation pump is used to circulate the gas through a cold getter (SAES MicroTorr HP190-902F). After reaching the desired xenon pressure, monitored using a pressure gauge installed in the system, the ion source is gradually heated by increasing the filament current to 1.7 A. Once this current is achieved, the gas recirculation system is activated, and the system is allowed to stabilize for approximately 30 minutes. This stabilization period also serves to purify the xenon gas and prevent contamination from residuals released by the ion source. 
\begin{figure}[t]
\begin{centering}
\includegraphics[width=1\columnwidth]{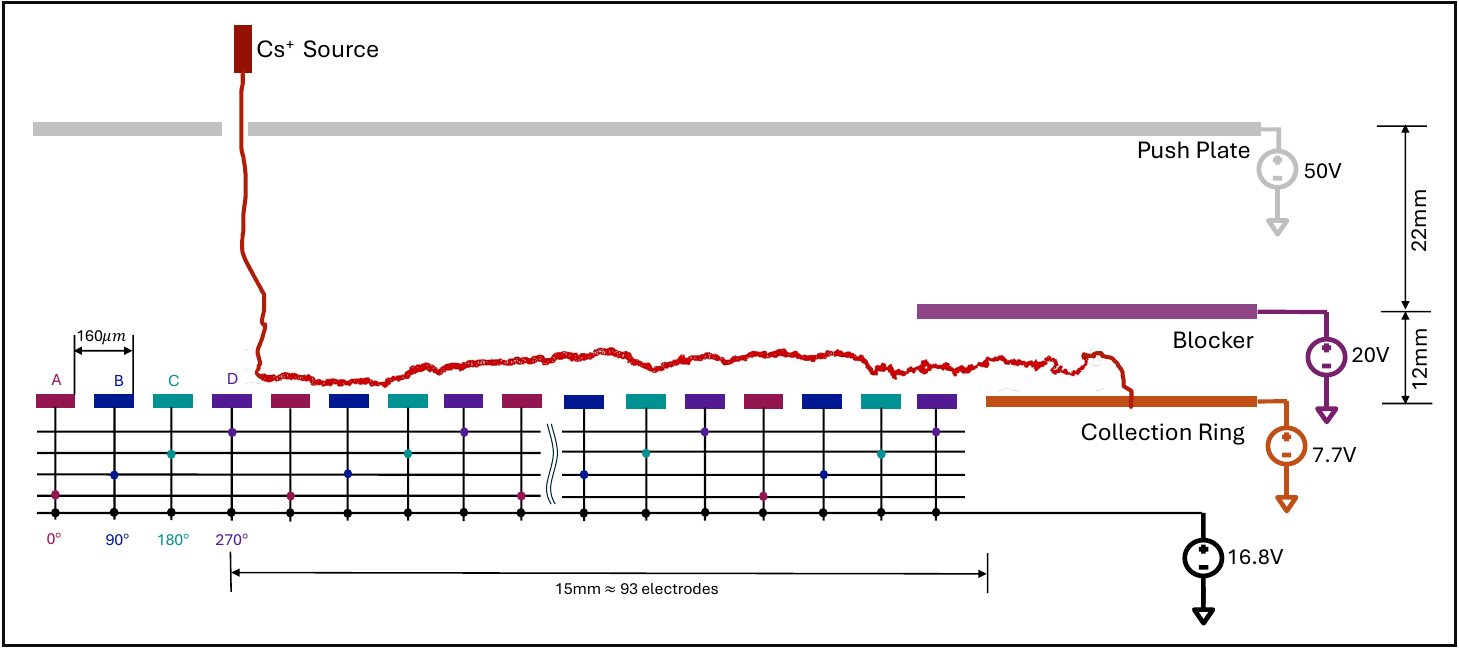}
\par\end{centering}
\caption{ Schematic of the four-phased RF carpet system including all the electrodes used.\label{fig:rfc}}
\end{figure}
Following stabilization, all DC electric fields are turned on, and the biasing of all components, except the ion source, is kept constant throughout the experiment. The voltages used in the different electrodes are shown in Fig.~\ref{fig:rfc}. Ion currents are measured at three distinct locations using picoammeters: approximately 40~nA on the push plate, 20~pA on the RF carpet, and 15–20~pA on the collection ring. For the measurement in the collection ring, a triaxial picoammeter was used to avoid noise at the picoampere level.

Ion transport efficiency is defined as the fraction of ions arriving at the carpet that are successfully swept to the collection ring. To measure the total ion current collected by the carpet ($I_{\text{C}}$), all electrodes on the carpet are shorted together, effectively treating the carpet as a single planar electrode. Afterward, the currents on the push plate ($I_P$) and the collection ring ($I$) are measured simultaneously, while the RF voltage $(V_{\text{RF}})$ is increased and keeping the push field constant. The current collected on the push plate serves as a calibration benchmark for the ion source output to monitor its stability. Due to source fluctuations, the number of ions reaching the carpet can vary slightly over time, but the number of ions reaching the carpet is linearly correlated to $I_P$. This means that $I_P$ can be used as a correction factor for $I_{\text{C}}$ when the measurements are taken. The bias applied to the ion source is also adjusted based on the ion current measured on the carpet, ensuring a consistent ion current of approximately 20~pA. The transport efficiency is calculated as:

\begin{equation}
    \epsilon = \frac{I(V_{\text{RF}}=V_{\text{pp}}) - I(V_{\text{RF}}=0)}{I_{\text{C}}\cdot I_P(V_{\text{RF}}=V_{\text{pp}})/{I_P(V_{\text{RF}}=0)}}\times 100 \%.
    \label{eq:6}
\end{equation}

\section{Results}\label{sec:result}

\subsection{Ion transport efficiencies}\label{sec:Ion}
\begin{figure}[t]
\begin{centering}
\includegraphics[width=1\columnwidth]{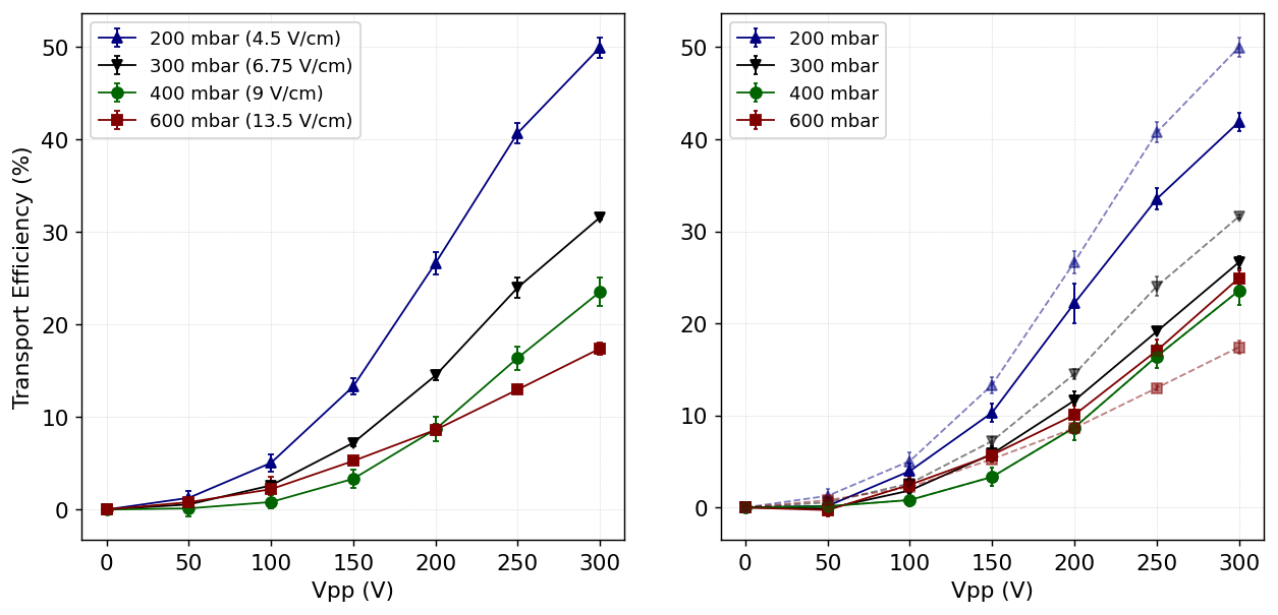}
\par\end{centering}
\caption{(Left) Transport efficiency of Cs$^{+}$ ions as a function of RF peak-to-peak voltage at various xenon gas pressures, using a scaled push field at each pressure. The push field values ($E_{push}$) applied at each pressure are indicated in parentheses. (Right) Comparison of transport efficiency as a function of xenon gas pressure, showing both fixed and scaled push field configurations. The solid lines correspond to a fixed push field of 9~V/cm (the value at 400~mbar), while the dashed lines represent the data from left plot (scaled push fields), overlaid for direct comparison.\label{fig:Efficiency}}
\end{figure}

Our benchmark operating point for the RF carpet described in this paper is transport of Cs$^+$ ions in 400~mbar of xenon gas. The RF carpet operates at a resonance frequency of 2.09~MHz with a push field of 9~V/cm (at 400~mbar), determined empirically through iterative adjustments during the experiment. Transport efficiency is measured as a function of RF amplitude. As shown in Fig.~\ref{fig:Efficiency}, the efficiency increases with RF voltage. However, the maximum RF voltage is limited by the breakdown strength of the carpet’s insulating material, polyimide. 

\begin{figure}[!ht]
\begin{centering}
\includegraphics[width=1\columnwidth]{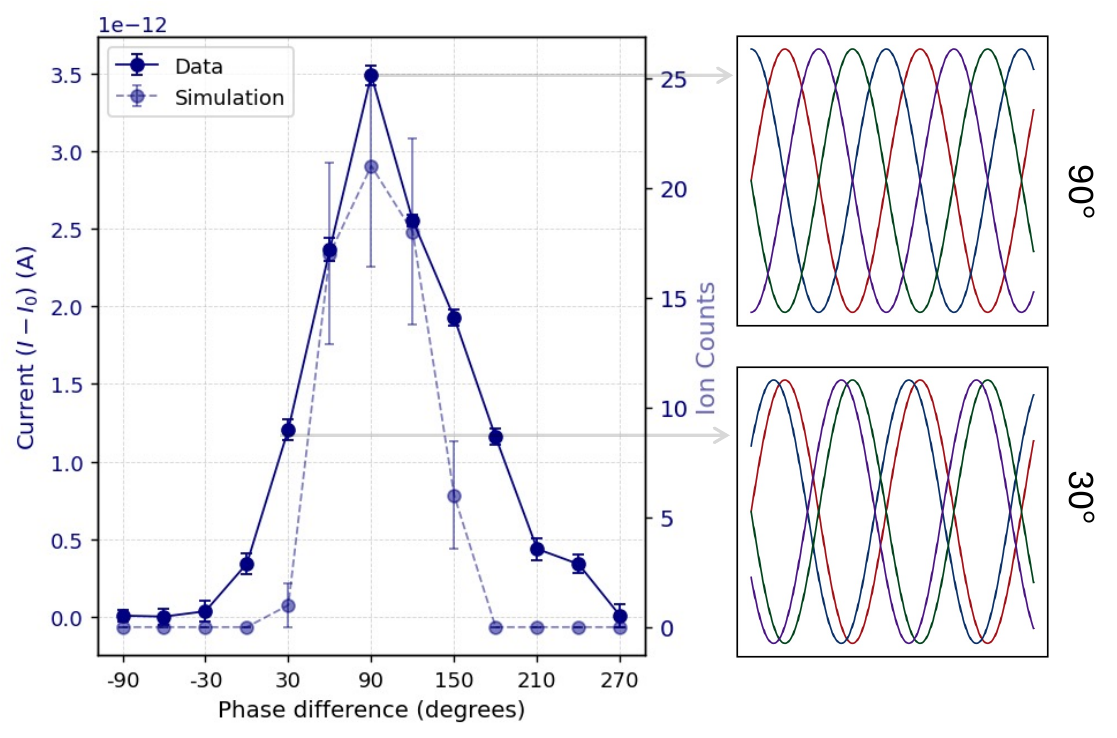}
\par\end{centering}
\caption{Phase plot demonstrating ion transport across the four-phased RF carpet as a function of phase difference between adjacent electrodes. The subplots show how the intermediate phases looks as a function of time between the four electrodes. The y-axis label is given by $I-I_0 = I(V_{\text{RF}}=V_{\text{pp}}) - I(V_{\text{RF}}=0$).\label{fig:phase}}
\end{figure}

To verify that any observed current is indeed driven by the RF traveling wave, we analyzed transport efficiency as a function of phase difference between consecutive RF electrodes. Strong transport is expected at 90$^\circ$ phase differences, whereas out-of-phase electrodes should not provide the required ion sweeping dynamics. The phase plot (Fig.~\ref{fig:phase}) demonstrates ion transport across a full $360^\circ$. It is noted that the phase study was generated at an RF amplitude of V$_{\text{pp}}$ = 250~V, since, with a phase difference of $180^\circ$, the maximal voltage between electrodes is $V_{\text{max}} = V_{\text{pp}}$, exceeding the value expected in the four-phase configuration ($V_{\text{max}} = 212$~V  at $V_{\text{pp}} = 300$~V).

The x-axis of the phase plot (Fig.~\ref{fig:phase}) shows the phase difference between electrode A and electrode B. When electrode A is ahead of electrode B by $90^\circ$, the ions are pushed radially outward. Conversely, when the phase is reversed, ions are transported toward the central hole. Since our collection region was located in the outer ring, maximum ion transport occurred when the phase difference was $90^\circ$, and no transport observed when ions were traveling in the opposite direction. Thus, the phase plot confirms that ion transport is primarily driven by the RF traveling wave and not by any DC or diffusive effects. 

The various DC voltages in the system were tuned to maximize transport at $90^\circ$ phase and in 400~mbar. We then examined ion transport efficiency at varying pressures. Since ion drift in DC fields primarily depends on the reduced electric field $(E/P)$, it is natural to scale the push plate voltages in this manner to facilitate comparison between pressures. However, caution should be observed in over-interpreting this kind of scaling, since it is not valid under conditions where RF dynamics are dominant. Since our study involves both ion drift to the carpet and then competition between RF and DC dynamics, there is no unique pressure/field scaling law expected. We thus consider two configurations.  In the study presented in Fig.~\ref{fig:Efficiency} (left) transport efficiency is presented with pressure-scaled push fields (keeping $E_{push}/P$ constant) but unscaled RF fields. Another possibility is to maintain the push fields at their 400~mbar derived values.  Fig.~\ref{fig:Efficiency} (left) shows the optimal transport we were able to achieve at each pressure point using this method.

Fig.~\ref{fig:Efficiency} (right) shows the transport efficiency as a function of $V_{\text{pp}}$ keeping the push field fixed to 9~V/cm. At 600~mbar, higher efficiency is achieved compared to 400~mbar, as the relatively lower push field for 600~mbar doesn’t allow ions to come close enough to the carpet
surface to interact with the RF pseudo-potential and as a result they’re drifted by the DC field. At 200~mbar, however, the relatively higher push field causes ions to collide with the carpet electrodes, decreasing efficiency.

\subsection{Comparison against simulations}
\label{sec:sim}

We compared the measured transport efficiencies to the transport efficiency obtained from the simulation of ion trajectories through a model of our system in SIMION~\cite{SIMION}, a widely used ion simulation software designed to model the trajectories of ions in electric and magnetic fields. We employed the hard sphere collision model in SIMION to investigate ion transport efficiency under varying pressure conditions. In one study, a total of 100 ions were simulated in an RF field of 300 $V_{\text{pp}}$, with pressures ranging from 100~mbar to 600~mbar. This pressure range was chosen to explore ion transport efficiency across different conditions. The input parameters for the SIMION simulations included pressure, electric fields, ion cross-section, and ion mass. 

While the pressure and electric fields were predetermined from the experimental setup, the ion-buffer-gas cross-section and effective mass were not measured. Past studies have demonstrated that ions can form clusters with xenon gas~\cite{benitezmedina2014}, with the rate and extent of clustering depending on pressure and ion species. This effect significantly complicates predictions of ion transport in dense xenon. The collision cross-section is directly influenced by the mass of the ion system, which can vary as clustering occurs. While the clustering dynamics for monocations like Cs$^+$ are expected to be far less strong than those in Ba$^{2+}$, much about this clustering behavior remains unknown. Therefore, we compare our experimental data to three cases.

The simplest assumption considers un-clustered Cs$^+$ ions colliding with Xe atoms, where the geometric cross-section, $\Omega_G$, is calculated as:
\begin{equation} \Omega_G = \pi (R_{\text{Xe}} + R_{\text{Cs}})^2 
\label{eq:7}
\end{equation}
Here, $R_{\text{Xe}}$ and $R_{\text{Cs}}$ represent the Van der Waals radii of xenon and caesium, with values $R_{\text{Xe}} = 216$~pm and $R_{\text{Cs}} = 343$~pm, respectively. This yields a geometric cross-section of $\Omega_G = 9.817 \times 10^{-15}$~cm$^2$.

\begin{figure}[t]
\begin{centering}
\includegraphics[width=1\columnwidth]{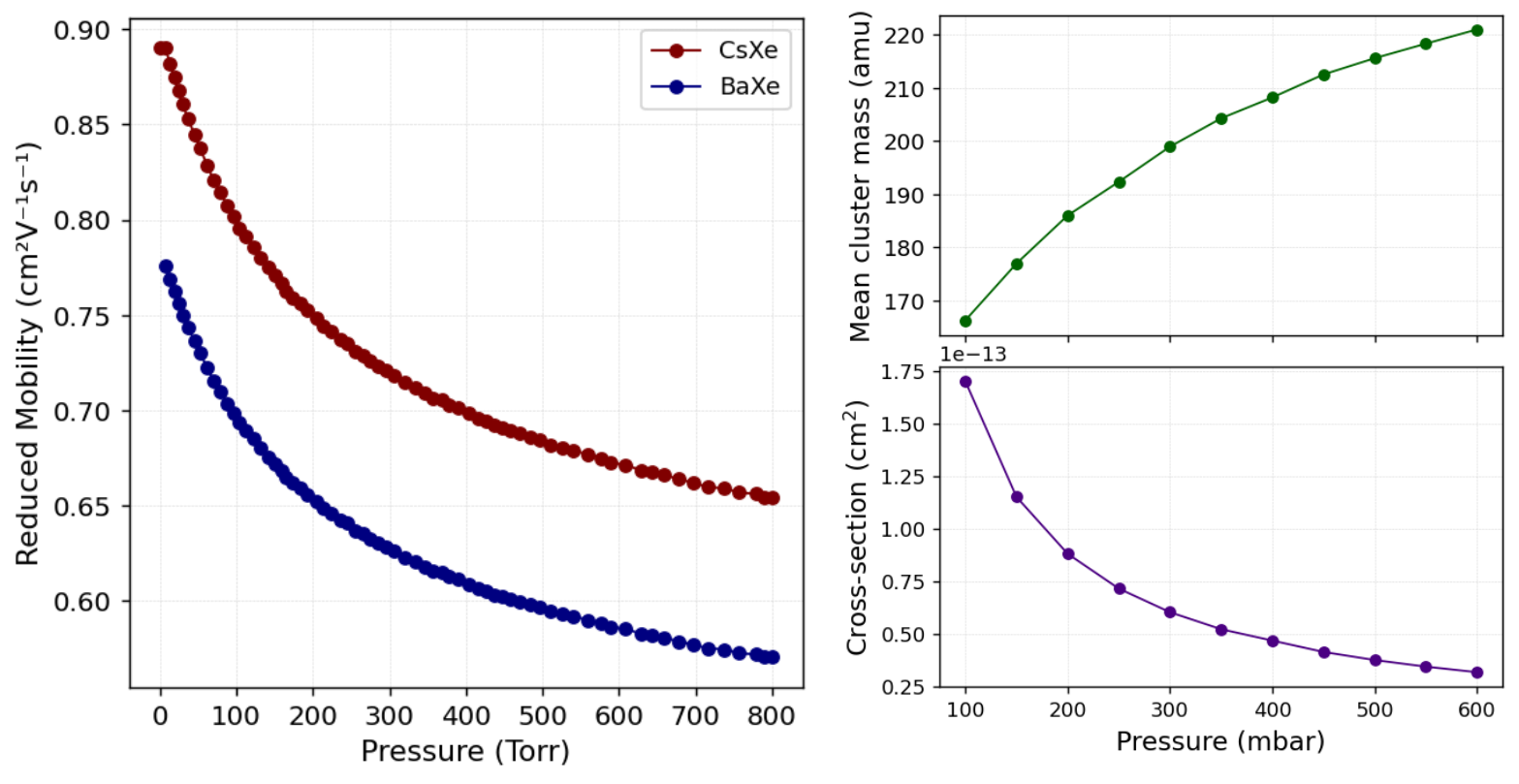}
\par\end{centering}
\caption{(Left) Reduced mobility of Cs$^{+}$/Xe and Ba$^{+}$/Xe as a function of pressure. At low pressures (near 0 Torr), the red data points represent experimentally measured Cs$^{+}$/Xe mobility. Beyond this low-pressure region, the red curve is derived by scaling the measured Cs$^{+}$/Xe mobility according to the fractional reduction observed in Ba$^{+}$/Xe data from Ref.~[26]. The blue data points show the Ba$^{+}$/Xe reference for comparison.
(Top Right) Dependence of the mean cluster mass on pressure for the Cs$^{+}$/Xe system under the assumed clustering model.
(Bottom Right) Scaled cross-section values calculated using the pressure-dependent scaled Cs$^{+}$/Xe mobility.
\label{fig:mobility}}
\end{figure}

\begin{figure}[t]
\begin{centering}
\includegraphics[width=0.7\columnwidth]{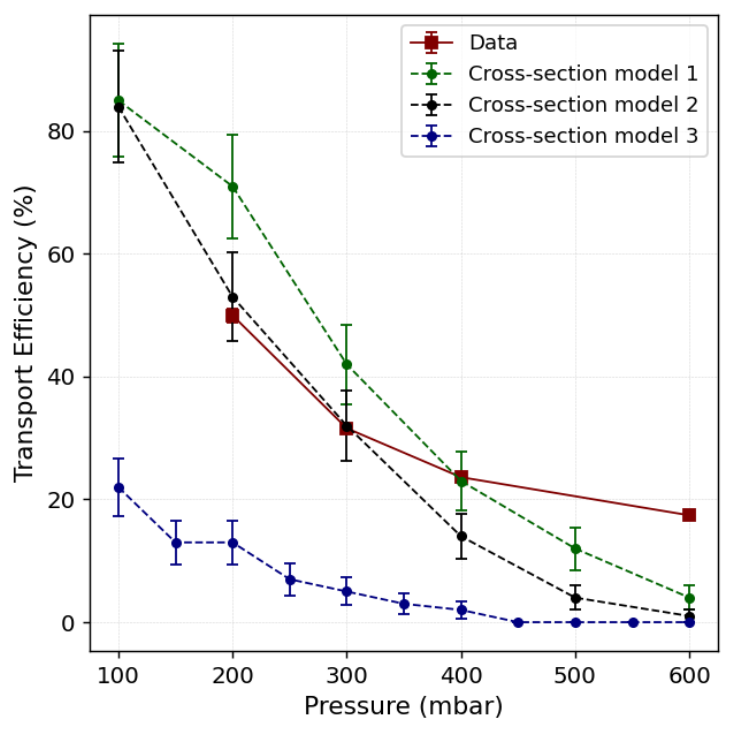}
\par\end{centering}
\caption{Ion transport efficiency as a function of pressure for $V_{\mathrm{pp}}=300\,\mathrm{V}$. 
    The red data points correspond to the experimentally measured values from Fig.~\ref{fig:Efficiency} (left). 
    The remaining curves show SIMION simulation results derived using three different cross-section models:
    (1) a purely geometric cross-section,
    (2) a cross-section calculated without accounting for ion clustering, and
    (3) a pressure-dependent cross-section inferred from the clustering model shown in Fig.~\ref{fig:mobility} (bottom right).}\label{fig:SIMION_sim}
\end{figure}

Another possibility is to determine the cross-section semi-empirically. Under the hard-sphere collision dynamics model there is an analytical expression relating momentum transfer cross-section $\Omega_M$ to the experimental mobility $\mu$,
\begin{equation}
\Omega_M = \frac{3q}{16n} \sqrt{\frac{2\pi}{\mathcal{M}kT}} \frac{1}{\mu}.
\label{eq:8}
\end{equation}
Here, $n$ is the gas number density, $k$ is the Boltzmann constant, $T$ is the temperature, of the ion in the buffer gas and $\mathcal{M}$ is the reduced mass,
\begin{equation}
\mathcal{M} = \frac{mM}{m + M}.
\label{eq:9}
\end{equation}
The reduced mobility (normalized ion mobility for a pressure P and temperature T, $\mu_0 = \mu \cdot \frac{P}{P_0} \frac{T_0}{T}$, where $P_0$ and $T_0$ being the standard temperature and pressure) of Cs${^+}$ in Xe has been measured at low pressure (0.03 - 0.6 mbar), and is reported in~\cite{thackston1978tests, viehland2017accurate}, with a value of $\mu_0$ = 0.9018~cm$^2 /$Vs when scaled to a temperature of 300~K and a pressure of 1.013~bar. Using $M = 131$~amu (the mass of xenon) and $m = 133$~amu (the mass of caesium) to calculate the reduced mass, we obtain the momentum transfer cross-section as  $\Omega_M = 1.603 \times 10^{-14}$~cm$^2$. Note that, we have neglected any clustering effect while calculating the reduced mass and subsequently the momentum transfer cross-section.

The mobility of Cs$^{+}$ ions in xenon at high pressures remains poorly characterized, and existing experimental values do not account for clustering effects that can significantly alter ion mobility at elevated pressures. Previous density functional theory-based simulations~\cite{Bainglass:2018odn} for Ba$^{+}$/Xe have successfully incorporated ion clustering and reproduced experimental observations. Drawing on this analogy, we can estimate Cs$^{+}$/Xe clustering behavior using the Ba$^{+}$/Xe system as a reference. By applying the same fractional reduced mobility scaling observed in Ba$^{+}$/Xe to Cs$^{+}$/Xe, we obtain the pressure-dependent mobility scaling shown in Fig.~\ref{fig:mobility} (left). This model begins with the experimentally measured reduced mobility of Cs$^{+}$ at very low pressures and then adjusts it to account for ion clustering effects as pressure increases.

From the Ba$^{+}$/Xe system, one can determine the relative fractions of Ba$^{+}$ and [BaXe]$^{+}$ ions (i.e., the fraction of Ba$^{+}$ and [BaXe]$^{+}$ ions out of the total ion population) present. Assuming that the Cs$^{+}$/Xe system exhibits similar pressure-dependent clustering dynamics, we adopt the same fraction of Cs$^{+}$ and [CsXe]$^{+}$ ions as inferred for Ba$^{+}$/Xe. In doing so, the mean cluster mass at each pressure is given by
\begin{equation} (f_{\text{Cs}} m_{\text{Cs}} + f_{\text{CsXe}} m_{\text{CsXe}}) \end{equation} where $f_{\text{Cs}}$ and $f_{\text{CsXe}}$ are the ion fractions and $m_{\text{CsXe}}$ = 264~amu is the mass of a [CsXe]$^{+}$ ion. The resulting mean cluster mass as a function of pressure is depicted in Fig. \ref{fig:mobility} (top right). Using this mean cluster mass, together with the scaled mobility, we recalculate the scattering cross-section according to Eq. \eqref{eq:8}. The calculated cross section in this approach is shown in Fig. \ref{fig:mobility} (bottom right).

Fig.~\ref{fig:SIMION_sim} compares various theoretical estimates of transport efficiency to the experimental data. While the theoretical models clearly capture the overall downward trend in efficiency with increasing pressure and are broadly predictive of the pressure range over which this transition occurs, they do not provide a quantitatively accurate match to the measured values. The remaining discrepancies could stem from an incomplete treatment of ion clustering and associated RF transport phenomena, or from more complex experimental factors such as space charge and diffusive effects that are not fully accounted for in our current calculations. To achieve a more quantitative description of ion transport in this regime, further experimental and theoretical investigations will both be necessary.  Nevertheless, this experimental confirmation of the general trends predicted by the theoretical calculations in the pressure range of interest is an important validation of the viability of RF ion transport methods for use in dense gases.

\section{Conclusions and Outlook}
\label{sec:conc}

In this work, we have demonstrated, for the first time, the efficient lateral transport of heavy ions such as Cs$^+$ in moderate-pressure xenon gas using a four-phased RF carpet. This important proof of principle has shown RF transport of ions with significant efficiency at pressures up to 600~mbar, the highest pressure use of ion-sweeping RF carpets ever to our knowledge. Transport performance is limited by the maximum RF voltage that can be applied given breakdown strength of the insulating material of the carpet. 

The performance is in qualitative agreement with simulations made using the SIMION package. Quantitative predictability is limited by, among other issues, the phenomenon of ion clustering, particularly the formation of [CsXe]$^+$ clusters at elevated pressures. This clustering affects transport properties of the ions through their mobility and collision cross-sections. While further theoretical work is needed to understand this behaviour in the Cs$^+$/Xe system, it is notable that the experimental data show an efficiency exceeding the predictions at the highest pressures.
 
To achieve high transport efficiencies at the operating pressures needed for the barium tagging application, ongoing research is focused on reducing the pitch size of the RF carpet, which could enable transport efficiency in 5-10~bar environments. The first prototypes of new generation of RF carpets have already been fabricated via photolithography specifically for the NEXT application. These devices will operate with pitches as low as 20~$\mu$m. %Applying our qualitatively validated simulation methods suggest that ion transport in high-pressure gaseous xenon as a working medium may be achievable with such devices. 
While simple ion-tracking simulations may require refinement at higher pressures to achieve accurate quantitative results, their overall qualitative validation suggests that ion transport in high-pressure xenon gas will be feasible with finer-pitch devices, guiding future efforts to improve both modeling and device performance. This workrepresents an important step towards RF carpets as an integral element in an ion collecting and imaging scheme for future background-free neutrinoless double beta decay experiments.

\acknowledgments

The work in this paper was funded by the US Department of Energy under DE-SC0019054 and  DE-SC0019223 to the University of Texas at Arlington. The NEXT Collaboration acknowledges support from the following agencies and institutions: the European Research Council (ERC) under Grant Agreement No. 951281-BOLD; the European Union’s Framework Programme for Research and Innovation Horizon 2020 (2014–2020) under Grant Agreement No. 957202-HIDDEN; the MCIN/AEI of Spain and ERDF A way of making Europe under grants PID2021-125475NB and RTI2018-095979, and the Severo Ochoa and Mar\'ia de Maeztu Program grants CEX2023-001292-S, CEX2023-001318-M and CEX2018-000867-S; the Generalitat Valenciana of Spain under grants PROMETEO/2021/087 and CISEJI/2023/27; the Department of Education of the Basque Government of Spain under the predoctoral training program non-doctoral research personnel; the Spanish la Caixa Foundation (ID 100010434) under fellowship code LCF/BQ/PI22/11910019; the Portuguese FCT under project UID/FIS/04559/2020 to fund the activities of LIBPhys-UC; the Israel Science Foundation (ISF) under grant 1223/21; the Pazy Foundation (Israel) under grants 310/22, 315/19 and 465; the US Department of Energy under contracts number DE-AC02-06CH11357 (Argonne National Laboratory), DE-AC02-07CH11359 (Fermi National Accelerator Laboratory), DE-FG02-13ER42020 (Texas A\&M), DE-SC0019054 (Texas Arlington) and DE-SC0019223 (Texas Arlington); the US National Science Foundation under award number NSF CHE 2004111; the Robert A Welch Foundation under award number Y-2031-20200401; Deutsche Forschungsgemeinschaft (DFG, German Research Foundation) - AY 155/2-1; MB acknowledge support from the National Science Foundation under grant number PHY-2310059. Finally, we are grateful to the Laboratorio Subterr\'aneo de Canfranc for hosting and supporting the NEXT experiment.

\bibliographystyle{JHEP}
\bibliography{RFC}

\providecommand{\href}[2]{#2}\begingroup\raggedright\begin{thebibliography}{10}

\bibitem{Mass_spec1997}
S.A.~Shaffer, K.~Tang, G.A.~Anderson, D.C.~Prior, H.R.~Udseth and R.D.~Smith, \emph{A novel ion funnel for focusing ions at elevated pressure using electrospray ionization mass spectrometry}, \href{https://doi.org/10.1002/(SICI)1097-0231(19971030)11:16<1813::AID-RCM87>3.0.CO;2-D}{\emph{Rapid Communications in Mass Spectrometry} {\bfseries 11} (1997) 1813}.

\bibitem{Masssepec2}
{Anthony D. Appelhans, David A. Dahl}, \emph{The ion funnel: Theory, implementations, and applications}, {\emph{Mass Spectrometry Reviews} {\bfseries 29} (2010) 294}.

\bibitem{Riken}
P.~Schury, M.~Wada, Y.~Ito, F.~Arai, D.~Kaji, S.~Kimura et~al., \emph{Status of the low-energy super-heavy element facility at riken}, \href{https://doi.org/10.1016/j.nimb.2016.02.061}{\emph{Nuclear Instruments and Methods in Physics Research Section B: Beam Interactions with Materials and Atoms} {\bfseries 376} (2016) 425}.

\bibitem{Caribu}
G.~Savard, S.~Baker, C.~Davids, A.~Levand, E.~Moore, R.~Pardo et~al., \emph{Radioactive beams from gas catchers: The caribu facility}, \href{https://doi.org/10.1016/j.nimb.2008.05.091}{\emph{Nuclear Instruments and Methods in Physics Research Section B: Beam Interactions with Materials and Atoms} {\bfseries 266} (2008) 4086}.

\bibitem{brown1991quantum}
L.S.~Brown, \emph{Quantum motion in a paul trap}, \href{https://doi.org/10.1103/PhysRevLett.66.527}{\emph{Phys. Rev. Lett.} {\bfseries 66} (1991) 527}.

\bibitem{horvath1997fundamental}
G.~Horvath, R.~Thompson and P.~Knight, \emph{Fundamental physics with trapped ions}, \href{https://doi.org/10.1080/001075197182540}{\emph{Contemporary Physics} {\bfseries 38} (1997) 25}.

\bibitem{mehta2016integrated}
K.~Mehta, C.~Bruzewicz, R.~McConnell, R.~Ram, J.~Sage and J.~Chiaverini, \emph{Integrated optical addressing of an ion qubit}, \href{https://doi.org/10.1038/nnano.2016.139}{\emph{Nature Nanotechnology} {\bfseries 11} (2016) 1066}.

\bibitem{wang2017single}
Y.~Wang, M.~Um, J.~Zhang, S.~An, M.~Lyu, J.-N.~Zhang et~al., \emph{Single-qubit quantum memory exceeding ten-minute coherence time}, \href{https://doi.org/10.1038/s41566-017-0007-1}{\emph{Nature Photonics} {\bfseries 11} (2017) 646}.

\bibitem{brunner2015rf}
T.~Brunner, D.~Fudenberg, V.~Varentsov, A.~Sabourov, G.~Gratta, J.~Dilling et~al., \emph{An rf-only ion-funnel for extraction from high-pressure gases}, \href{https://doi.org/10.1016/j.ijms.2015.01.003}{\emph{International Journal of Mass Spectrometry} {\bfseries 379} (2015) 110}.

\bibitem{jones2022dynamics}
B.~Jones, A.~Raymond, K.~Woodruff, N.~Byrnes, A.~Denisenko, F.~Foss et~al., \emph{The dynamics of ions on phased radio-frequency carpets in high pressure gases and application for barium tagging in xenon gas time projection chambers}, \href{https://doi.org/10.1016/j.nima.2022.167000}{\emph{Nuclear Instruments and Methods in Physics Research Section A: Accelerators, Spectrometers, Detectors and Associated Equipment} {\bfseries 1039} (2022) 167000}.

\bibitem{buchmuller2005leptogenesis}
W.~Buchmuller, R.~Peccei and T.~Yanagida, \emph{Leptogenesis as the origin of matter}, \href{https://doi.org/10.1146/annurev.nucl.55.090704.151558}{\emph{Annual Review of Nuclear and Particle Science} {\bfseries 55} (2005) 311}.

\bibitem{dolinski2019neutrinoless}
M.J.~Dolinski, A.W.P.~Poon and W.~Rodejohann, \emph{{Neutrinoless Double-Beta Decay: Status and Prospects}}, \href{https://doi.org/10.1146/annurev-nucl-101918-023407}{\emph{Ann. Rev. Nucl. Part. Sci.} {\bfseries 69} (2019) 219} [\href{https://arxiv.org/abs/1902.04097}{{\ttfamily 1902.04097}}].

\bibitem{agostini2017discovery}
M.~Agostini, G.~Benato and J.A.~Detwiler, \emph{Discovery probability of next-generation neutrinoless double-$\ensuremath{\beta}$ decay experiments}, \href{https://doi.org/10.1103/PhysRevD.96.053001}{\emph{Phys. Rev. D} {\bfseries 96} (2017) 053001}.

\bibitem{Moe:1991ik}
M.K.~Moe, \emph{{New approach to the detection of neutrinoless double beta decay}}, \href{https://doi.org/10.1103/PhysRevC.44.931}{\emph{Physical Review} {\bfseries C44} (1991) 931}.

\bibitem{NEXT:2020amj}
C.~Adams et~al., \emph{Sensitivity of a tonne-scale next detector for neutrinoless double beta decay searches}, \href{https://doi.org/10.1007/JHEP08(2021)164}{\emph{JHEP} {\bfseries 2021} (2021) 164}.

\bibitem{Chambers:2018srx}
{\scshape nEXO} collaboration, \emph{{Imaging individual barium atoms in solid xenon for barium tagging in nEXO}}, \href{https://doi.org/10.1038/s41586-019-1169-4}{\emph{Nature} {\bfseries 569} (2019) 203} [\href{https://arxiv.org/abs/1806.10694}{{\ttfamily 1806.10694}}].

\bibitem{mong:2014iya}
B.~Mong, S.~Cook, T.~Walton, C.~Chambers, A.~Craycraft, C.~Benitez-Medina et~al., \emph{{Spectroscopy of Ba and Ba$^+$ deposits in solid xenon for barium tagging in nEXO}}, \href{https://doi.org/10.1103/PhysRevA.91.022505}{\emph{Physical Review} {\bfseries A91} (2015) 022505} [\href{https://arxiv.org/abs/1410.2624}{{\ttfamily 1410.2624}}].

\bibitem{rollin:2011gla}
E.~Rollin, \emph{{Barium Ion Extraction and Identification from Laser Induced Fluorescence in Gas for the Enriched Xenon Observatory}}, Ph.D. thesis, Carleton U., 2011.

\bibitem{sinclair:2011zz}
D.~Sinclair, E.~Rollin, J.~Smith, A.~Mommers, N.~Ackerman, B.~Aharmim et~al., \emph{{Prospects for Barium Tagging in Gaseous Xenon}}, \href{https://doi.org/10.1088/1742-6596/309/1/012005}{\emph{Journal of Physics Conference Series} {\bfseries 309} (2011) 012005}.

\bibitem{flatt:2007aa}
B.~Flatt, M.~Green, J.~Wodin, R.~DeVoe, P.~Fierlinger, G.~Gratta et~al., \emph{{A linear RFQ ion trap for the Enriched Xenon Observatory}}, \href{https://doi.org/10.1016/j.nima.2007.05.123}{\emph{Nuclear Instruments and Methods in Physics Research Section A: Accelerators, Spectrometers, Detectors and Associated Equipment} {\bfseries A578} (2007) 399} [\href{https://arxiv.org/abs/0704.1646}{{\ttfamily 0704.1646}}].

\bibitem{mcdonald2018demonstration}
{\scshape NEXT} collaboration, \emph{Demonstration of single-barium-ion sensitivity for neutrinoless double-beta decay using single-molecule fluorescence imaging}, \href{https://doi.org/10.1103/PhysRevLett.120.132504}{\emph{Phys. Rev. Lett.} {\bfseries 120} (2018) 132504}.

\bibitem{rivilla2020fluorescent}
I.~Rivilla, B.~Aparicio, J.~Bueno, D.~Casanova, C.~Tonnellé, Z.~Freixa et~al., \emph{Fluorescent bicolour sensor for low-background neutrinoless double $\beta$ decay experiments}, \href{https://doi.org/10.1038/s41586-020-2431-5}{\emph{Nature} {\bfseries 583} (2020) 48}.

\bibitem{jones2016single}
B.~Jones, A.~McDonald and D.~Nygren, \emph{Single molecule fluorescence imaging as a technique for barium tagging in neutrinoless double beta decay}, {\emph{Journal of Instrumentation} {\bfseries 11} (2016) P12011}.

\bibitem{thapa2021demonstration}
P.~Thapa, N.K.~Byrnes, A.A.~Denisenko, J.X.~Mao, A.D.~McDonald, C.A.~Newhouse et~al., \emph{Selective single-barium ion detection with dry diazacrown ether naphthalimide turn-on chemosensors}, \href{https://doi.org/10.1021/acssensors.0c02104}{\emph{ACS Sensors} {\bfseries 6} (2021) 192}.

\bibitem{thapa2019barium}
P.~Thapa, I.~Arnquist, N.~Byrnes, A.~Denisenko, F.W.F.~Jr. and B.J.P.~Jones, \emph{Barium chemosensors with dry-phase fluorescence for neutrinoless double beta decay}, \href{https://doi.org/10.1038/s41598-019-49283-x}{\emph{Scientific Reports} {\bfseries 9} (2019) 15097}.

\bibitem{Bainglass:2018odn}
E.~Bainglass, B.P.~Jones, F.~Foss, M.~Huda and D.~Nygren, \emph{{Mobility and Clustering of Barium Ions and Dications in High Pressure Xenon Gas}}, \href{https://doi.org/10.1103/PhysRevA.97.062509}{\emph{Phys. Rev. A} {\bfseries 97} (2018) 062509} [\href{https://arxiv.org/abs/1804.01169}{{\ttfamily 1804.01169}}].

\bibitem{herrero2022ba}
P.~Herrero-Gómez, J.~Calupitan, M.~Ilyn, A.~Berdonces-Layunta, T.~Wang, D.~de~Oteyza et~al., \emph{Ba$^{2+}$ ion trapping using organic submonolayer for ultra-low background neutrinoless double beta detector}, \href{https://doi.org/10.1038/s41467-022-35153-0}{\emph{Nature Communications} {\bfseries 13} (2022) 7741}.

\bibitem{nygren2015detecting}
D.R.~Nygren, \emph{Detecting the barium daughter in {$^{136}$Xe} 0-{$\nu\beta\beta$} decay using single-molecule fluorescence imaging techniques}, \href{https://doi.org/10.1088/1742-6596/650/1/012002}{\emph{Journal of Physics: Conference Series} {\bfseries 650} (2015) 012002}.

\bibitem{ray2024ion}
D.~Ray, R.~Collister, H.~Rasiwala, L.~Backes, A.V.~Balbuena, T.~Brunner et~al., \emph{Ion manipulation from liquid xe to vacuum: Ba-tagging for a nexo upgrade and future $0\nu\beta\beta$ experiments}, {\emph{arXiv preprint} (2024) } [\href{https://arxiv.org/abs/2410.18138}{{\ttfamily 2410.18138}}].

\bibitem{Microscope}
N.~Byrnes, E.~Dey, F.~Foss, B.~Jones, D.~Nygren et~al., \emph{Fluorescence imaging of individual ions and molecules in pressurized noble gases for barium tagging in {$^{136}$Xe}}, \href{https://doi.org/10.1038/s41467-024-54872-0}{\emph{Nature Communications} {\bfseries 15} (2024) 10595}.

\bibitem{Dehmelt}
H.~Dehmelt, \emph{Radiofrequency spectroscopy of stored ions i: Storage}, \href{https://doi.org/10.1016/S0065-2199(08)60170-0}{\emph{Advances in Atomic and Molecular Physics} {\bfseries 3} (1968) 53}.

\bibitem{Schwartz}
S.~Schwarz, \emph{Rf ion carpets: The electric field, the effective potential, operational parameters and an analysis of stability}, \href{https://doi.org/10.1016/j.ijms.2010.09.021}{\emph{International Journal of Mass Spectrometry} {\bfseries 299} (2011) 71}.

\bibitem{Bollen2011}
G.~Bollen, \emph{"ion surfing" with radiofrequency carpets}, \href{https://doi.org/10.1016/j.ijms.2010.09.032}{\emph{International Journal of Mass Spectrometry} {\bfseries 299} (2011) 131}.

\bibitem{MWada_circ_RFC}
F.~Arai, Y.~Ito, M.~Wada, P.~Schury, T.~Sonoda and H.~Mita, \emph{Investigation of the ion surfing transport method with a circular rf carpet}, \href{https://doi.org/10.1016/j.ijms.2014.01.005}{\emph{International Journal of Mass Spectrometry} {\bfseries 362} (2014) 56}.

\bibitem{Maxime2022RFC}
C.~Davis, R.~Bualuan, O.~Bruce, D.~Burdette, A.~Cannon, T.~Florenzo et~al., \emph{Commissioning of the st. benedict rf carpet}, \href{https://doi.org/10.1016/j.nima.2022.167422}{\emph{Nuclear Instruments and Methods in Physics Research Section A: Accelerators, Spectrometers, Detectors and Associated Equipment} {\bfseries 1042} (2022) 167422}.

\bibitem{RFC3}
{A.E. Gehring, M. Brodeur, G. Bollen, D.J. Morrissey, S. Schwarz}, \emph{Research and development of ion surfing rf carpets for the cyclotron gas stopper at the nscl.}, {\emph{Nuclear Instruments and Methods in Physics Research Section B: Beam Interactions with Materials and Atoms} {\bfseries 376} (2016) 221}.

\bibitem{RFC4}
L.~Querci, V.~Varentsov, D.~Günther and B.~Hattendorf, \emph{An rf-only ion funnel interface for ion cooling in laser ablation time of flight mass spectrometry}, \href{https://doi.org/10.1016/j.sab.2018.05.004}{\emph{Spectrochimica Acta Part B: Atomic Spectroscopy} {\bfseries 146} (2018) 57}.

\bibitem{RFC5}
M.~Ranjan, S.~Purushothaman, T.~Dickel, H.~Geissel, W.~Plaß, D.~Schäfer et~al., \emph{New stopping cell capabilities: Rf carpet performance at high gas density and cryogenic operation}, \href{https://doi.org/10.1209/0295-5075/96/52001}{\emph{Europhysics Letters (EPL)} {\bfseries 96} (2011) 52001}.

\bibitem{Lund2020}
K.~Lund, G.~Bollen, D.~Lawton, D.~Morrissey, J.~Ottarson, R.~Ringle et~al., \emph{Online tests of the advanced cryogenic gas stopper at nscl}, \href{https://doi.org/10.1016/j.nimb.2019.04.053}{\emph{Nuclear Instruments and Methods in Physics Research Section B: Beam Interactions with Materials and Atoms} {\bfseries 463} (2020) 378}.

\bibitem{ringle2021particle}
R.~Ringle, G.~Bollen, K.~Lund, C.~Nicoloff, S.~Schwarz, C.~Sumithrarachchi et~al., \emph{Particle-in-cell techniques for the study of space charge effects in the advanced cryogenic gas stopper}, \href{https://doi.org/10.1016/j.nimb.2021.03.020}{\emph{Nuclear Instruments and Methods in Physics Research Section B: Beam Interactions with Materials and Atoms} {\bfseries 496} (2021) 61}.

\bibitem{savard2016caribu}
G.~Savard, A.~Levand and B.~Zabransky, \emph{The caribu gas catcher}, \href{https://doi.org/10.1016/j.nimb.2016.02.050}{\emph{Nuclear Instruments and Methods in Physics Research Section B: Beam Interactions with Materials and Atoms} {\bfseries 376} (2016) 246}.

\bibitem{St.Benedict}
C.~Davis, O.~Bruce, D.~Burdette, T.~Florenzo, B.~Liu, J.~Long et~al., \emph{Transport tests of the st. benedict first-stage extraction system}, \href{https://doi.org/10.1016/j.nima.2022.166509}{\emph{Nuclear Instruments and Methods in Physics Research Section A: Accelerators, Spectrometers, Detectors and Associated Equipment} {\bfseries 1031} (2022) 166509}.

\bibitem{VNA}
\url{https://nanorfe.com/nanovna-v2.html}.

\bibitem{spice}
\url{https://www.analog.com/en/resources/design-tools-and-calculators/ltspice-simulator.html}.

\bibitem{Amplifier}
\url{https://www.dxworld-e.com/product-page/mrf300-ldmos-600w-hf-linear-amplifier-160-6-ldmos-included}.

\bibitem{Aluminosilicate_filament_source}
J.P.~Blewett and E.J.~Jones, \emph{Filament sources of positive ions}, \href{https://doi.org/10.1103/PhysRev.50.464}{\emph{Phys. Rev.} {\bfseries 50} (1936) 464}.

\bibitem{Weber_Cordes_1966}
R.E.~Weber and L.F.~Cordes, \emph{Aluminosilicate alkali ion sources}, \href{https://doi.org/10.1063/1.1719925}{\emph{Review of Scientific Instruments} {\bfseries 37} (1966) 112}.

\bibitem{Cathode_TechBulletin_118}
``Technical bulletin 118: Aluminosilicate cathodes.'' \url{https://www.cathode.com/pdf/tb-118.pdf}.

\bibitem{COMSOL}
E.J.~Dickinson, H.~Ekström and E.~Fontes, \emph{Comsol multiphysics®: Finite element software for electrochemical analysis. a mini-review}, \href{https://doi.org/10.1016/j.elecom.2013.12.020}{\emph{Electrochemistry Communications} {\bfseries 40} (2014) 71}.

\bibitem{SIMION}
A.D.~Appelhans and D.A.~Dahl, \emph{Simion ion optics simulations at atmospheric pressure}, \href{https://doi.org/10.1016/j.ijms.2005.03.010}{\emph{International Journal of Mass Spectrometry} {\bfseries 244} (2005) 1}.

\bibitem{benitezmedina2014}
J.C.B.~Medina, \emph{MOBILITY AND FLUORESCENCE OF BARIUM IONS IN XENON GAS FOR THE EXO EXPERIMENT}, Ph.D. thesis, Colorado State University, 2014.

\bibitem{thackston1978tests}
M.G.~Thackston, F.L.~Eisele, W.M.~Pope, H.W.~Ellis and E.W.~McDaniel, \emph{Further tests of the generalized einstein relation: Cs$^+$ ions in ar, kr, and xe}, \href{https://doi.org/10.1063/1.436208}{\emph{The Journal of Chemical Physics} {\bfseries 68} (1978) 3950}.

\bibitem{viehland2017accurate}
L.A.~Viehland, T.~Skaist, C.~Adhikari and W.F.~Siems, \emph{Accurate zero-field mobilities of atomic ions in the rare gases for calibration of ion mobility spectrometers}, \href{https://doi.org/10.1007/s12127-016-0212-5}{\emph{International Journal for Ion Mobility Spectrometry} {\bfseries 20} (2017) 1}.

\end{thebibliography}\endgroup

\end{document}